\documentclass[a4paper]{jpconf}
\usepackage{graphicx}
\begin{document}
\title{Early thermalization of quark-gluon matter initially created in
high-energy nucleus-nucleus collisions}

\author{Xiao-Ming Xu}

\address{Department of Physics, Shanghai University, Baoshan, Shanghai 200444, 
China}

\ead{xmxu@mail.shu.edu.cn}

\begin{abstract}
Elastic parton-parton-parton scattering is briefly reviewed and is included in
transport equations of quark-gluon matter. We solve the transport equations and
get thermal states from initially produced quark-gluon matter. Both gluon
matter and quark matter take early thermalization, but gloun matter has a
shorter thermalization time than quark matter.
\end{abstract}

\section{Introduction}
\medskip
A large amount of particles move randomly and momentum distribution functions
in different directions are identical. The distribution depends only on the
absolute value of the three-dimensional momentum. The dependence indicates
the existence of a thermal state and exposes the basic quantity temperature,
which value is acquired from the analytic expression of the distribution
(e.g., temperature is in the Boltzmann distribution).
Unfortunately, the system of
gluons, quarks and antiquarks produced in initial nucleus-nucleus collisions
at the Relativistic Heavy Ion Collider (RHIC) energies does not possess
temperature. This is because the distribution in the incoming nucleus beam
direction is much larger than the distribution in the direction perpendicular
to the beam direction \cite{xmxu1}. 
However, the system evolves rapidly into a thermal state (a quark-gluon plasma)
as concluded from the elliptic flow data of hadrons \cite{STAR} and the
corresponding explanation of hydrodynamic calculations which assume early
thermalization and ideal relativistic fluid flow \cite{kolb,teaney}.
The early thermalization or the rapid creation of temperature is generally
interesting. Parton-parton scattering was employed to study thermalization
of initially produced quark-gluon matter, but is not enough to explain the
early thermalization \cite{shuryak,wong,shin,blaizot}. 
Elastic gluon-gluon-gluon scattering was proposed to explain the
early thermalization of gluon matter \cite{xmxu2}. Nevertheless,
elastic quark-quark scattering and elastic quark-quark-quark scattering 
give a long thermalization time of quark matter \cite{xmxu3}. 
Recently we have obtained that quark matter takes the early thermalization 
due to all types of elastic scattering that involves quarks \cite{xmxu4}.
In the next section I briefly introduce the elastic scattering of
quark-quark-quark \cite{xmxu3}, quark-quark-antiquark \cite{xmxu5}, 
quark-antiquark-antiquark, antiquark-antiquark-antiquark, gluon-quark-quark 
\cite{xmxu6}, gluon-quark-antiquark \cite{xmxu6}, gluon-antiquark-antiquark,
gluon-gluon-quark \cite{xmxu4}, gluon-gluon-antiquark or gluon-gluon-gluon
\cite{xmxu2}. In section 3 we present transport equations that include all
types of elastic parton-parton-parton scattering and numerical results of the
equations. Conclusions are in the last section.

\begin{table}
\centering \caption{The number of diagrams for elastic parton-parton-parton
scattering, the
number of diagrams in a class, the number of the triple-gluon vertex and
the number of the four-gluon vertex of a diagram.}
\label{table1}
\begin{tabular*}{16cm}{@{\extracolsep{\fill}}c|c|c|c|c|c}

  \hline
  elastic & number of     & class & number of  & number of   & number of \\
  scattering & diagrams   &       & diagrams   & the 3-gluon & the 4-gluon \\
             & for a type &       & in a class & vertex      & vertex \\
  \hline
 quark-quark-quark 1      & 42   & 1   & 36   & 0  & 0 \\
                          &      & 2   & 6    & 1  & 0 \\
  \hline
 quark-quark-quark 2      & 14   & 1   & 12   & 0  & 0 \\
                          &      & 2   & 2    & 1  & 0 \\
  \hline
 quark-quark-quark 3      & 7    & 1   & 6    & 0  & 0 \\
                          &      & 2   & 1    & 1  & 0 \\
  \hline
 quark-quark-antiquark 1  & 58   & 1  & 52  & 0  & 0  \\
                          &      & 2  & 6   & 1  & 0  \\
  \hline
 quark-quark-antiquark 2  & 14   & 1  & 12  & 0  & 0  \\
                          &      & 2  & 2   & 1  & 0  \\
  \hline
 quark-quark-antiquark 3  & 29   & 1  & 26  & 0  & 0  \\
                          &      & 2  & 3   & 1  & 0  \\
  \hline
 quark-quark-antiquark 4  & 7    & 1  & 6   & 0  & 0  \\
                          &      & 2  & 1   & 1  & 0  \\
  \hline
 gluon-quark-quark 1    & 72  & 1  & 40  & 0     & 0  \\
                        &     & 2  & 24  & 1     & 0  \\
                        &     & 3  & 8   & 2     & 0  \\
                or      &     &    &     & 0     & 1  \\
  \hline
 gluon-quark-quark 2    & 36  & 1  & 20  & 0     & 0  \\
                        &     & 2  & 12  & 1     & 0  \\
                        &     & 3  & 4   & 2     & 0  \\
                or      &     &    &     & 0     & 1  \\
  \hline
 gluon-quark-antiquark 1    & 76  & 1  & 40  & 0     & 0  \\
                            &     & 2  & 28  & 1     & 0  \\
                            &     & 3  & 8   & 2     & 0  \\
                    or      &     &    &     & 0     & 1  \\
  \hline
 gluon-quark-antiquark 2    & 36  & 1  & 20  & 0     & 0  \\
                            &     & 2  & 12  & 1     & 0  \\
                            &     & 3  & 4   & 2     & 0  \\
                    or      &     &    &     & 0     & 1  \\
  \hline
 gluon-gluon-quark      & 123  & 1   & 24   & 0     & 0 \\
                        &      & 2   & 36   & 1     & 0 \\
                        &      & 3   & 38   & 2     & 0 \\
                or      &      &     &      & 0     & 1 \\
                        &      & 4   & 25   & 3     & 0 \\
                or      &      &     &      & 1     & 1 \\
  \hline
 gluon-gluon-gluon      & 220  & 1   & 105   & 4     & 0 \\
                        &      & 2   & 96    & 2     & 1 \\
                        &      & 3   & 19    & 0     & 2 \\
  \hline
 \end{tabular*}
\end{table}

\section{Elastic parton-parton-parton scattering}
\medskip
Each type of elastic 3-to-3 scattering is divided into 2, 3 or 4 classes
according to the number of the triple-gluon vertex and the number of the
four-gluon vertex. In Table 1 we list the number of diagrams for each type 
or in a class, the number
of the triple-gluon vertex and the number of the four-gluon vertex of a
diagram. The number following the entry quark-quark-quark has the meanings:
the three quarks are identical at 1; only two quarks are identical at 2; the
three quarks are different at 3. The number following the entry 
quark-quark-antiquark has the meanings: the two quarks and the antiquark 
have the same flavor at 1; only the two quarks have the same flavor at 2; only 
one quark has the same flavor as the antiquark at 3; the 
three flavors of the two
quarks and the antiquark are not identical at 4. The number following the entry
gluon-quark-quark (gluon-quark-antiquark) has the meanings: the two quarks
(the quark and the antiquark) have the same flavor at 1; the two quarks 
(the quark and the antiquark) take different flavors at 2. 

Feynman diagrams for the elastic quark-antiquark-antiquark scattering
(antiquark-antiquark-antiquark, gluon-antiquark-antiquark, 
gluon-gluon-antiquark) are obtained from the Feynman diagrams for the elastic
quark-quark-antiquark (quark-quark-quark, gluon-quark-quark, gluon-gluon-quark)
via the replacement of quark by antiquark and vice versa. To illustrate 
elastic 3-to-3 scattering, we show four Feynman diagrams in figures 1-4. 
In figures 1 and 2 quark-antiquark annihilation and creation occur. In both
figure 3 and figure 4 two triple-gluon vertices and one four-gluon vertex are
involved.

\begin{figure}
\begin{minipage}{16.5pc}
\includegraphics[width=12pc]{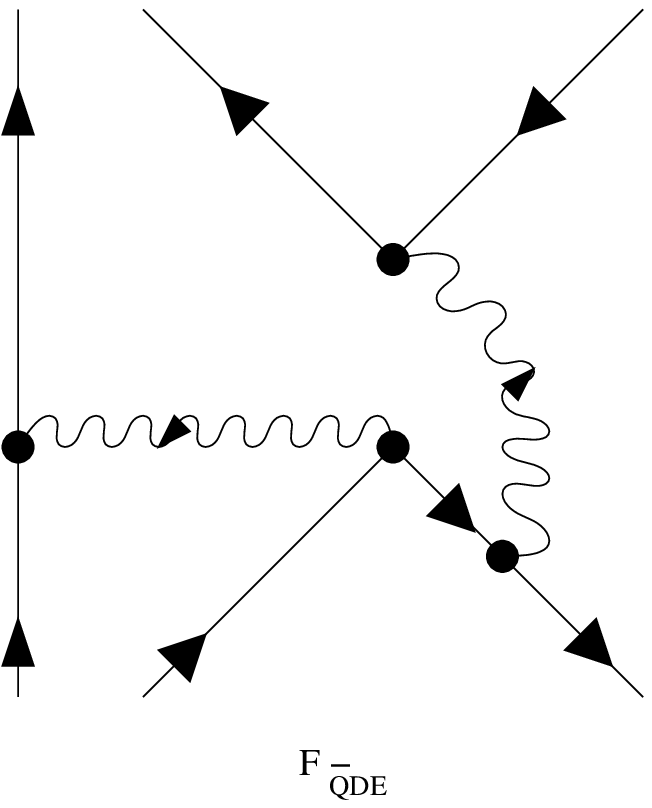}
\caption{\label{qqqbar1}Elastic gluon-quark-antiquark scattering.
The wiggly (solid) lines stand for gluons (quarks and antiquarks).}
\end{minipage}
\hspace{3pc}
\begin{minipage}{16.5pc}
\includegraphics[width=12pc]{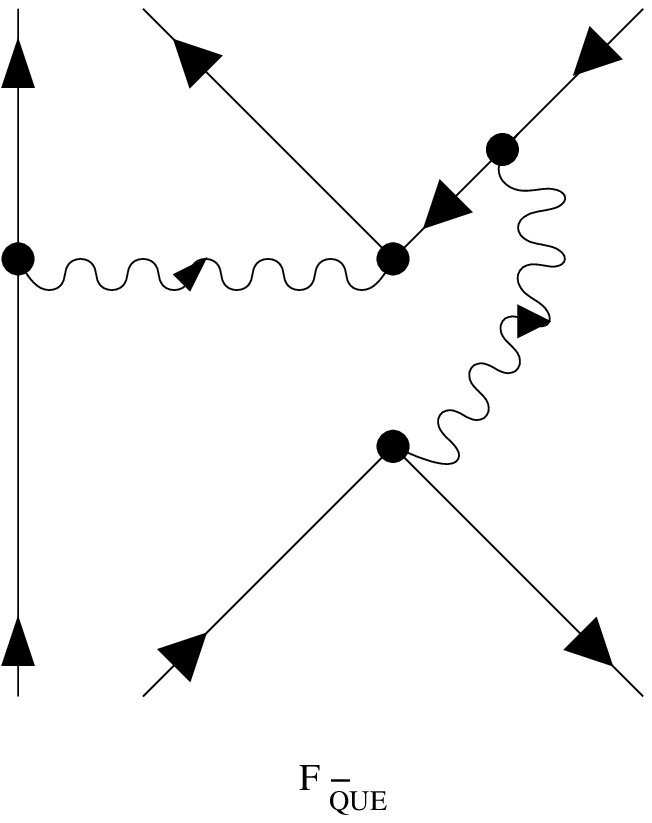}
\caption{\label{qqqbar2}Same as figure 1.}
\end{minipage}
\end{figure}

\begin{figure}
\begin{minipage}{16pc}
\includegraphics[width=10pc]{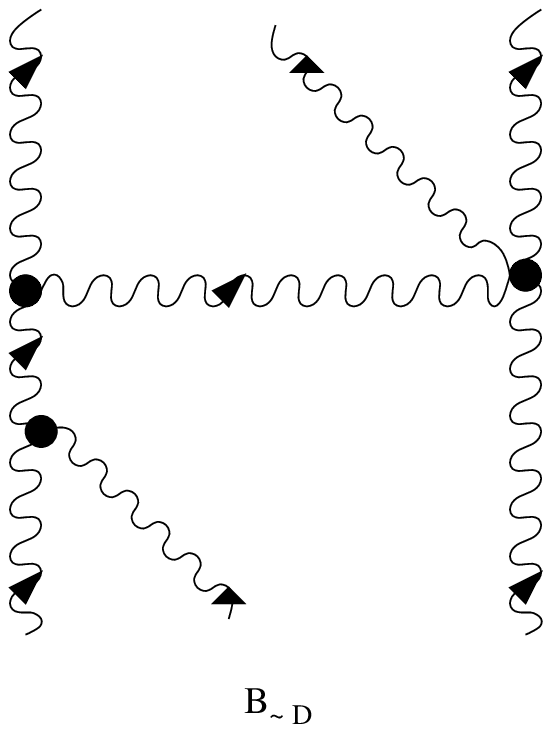}
\caption{\label{ggg1}Elastic gluon-gluon-gluon scattering.
The wiggly lines stand for gluons.}
\end{minipage}
\hspace{3pc}
\begin{minipage}{16pc}
\includegraphics[width=15pc]{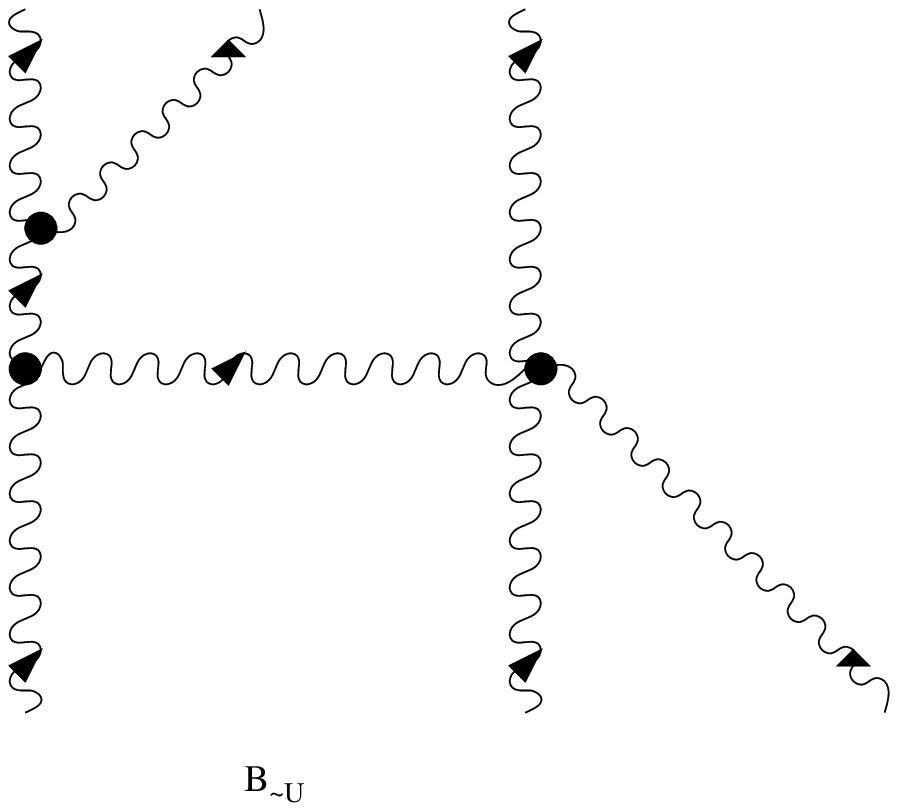}
\caption{\label{ggg2}Same as figure 3.}
\end{minipage}
\end{figure}

The elastic gluon-gluon-gluon scattering is complicated but rich in the process
as shown by the 220 Feynman diagrams at the tree level \cite{xmxu2}. 
To better understand
the scattering, we sort out Feynman diagrams in every class.
We select three triple-gluon vertices of a diagram to subclassify the diagrams
in the first class and six subclasses are obtained according to the 
initial gluons and the final gluons possessed by the selected vertices.
The subclasses are listed in Table 2. Every diagram in the second class has a
four-gluon vertex and two triple-gluon vertices.
We select the four-gluon vertex and a
triple-gluon vertex to subclassify the diagrams in the second
class and nine subclasses are obtained according to the initial
gluons and the final gluons possessed by the four-gluon vertex and the selected
triple-gluon vertex. The subclasses are listed in Table 3.
Diagrams in the third class can not be subclassified.

\begin{table}
\centering \caption{Number of diagrams in a subclass identified
by the three selected triple-gluon
vertices. $n_{1\rm i}$ and $n_{1\rm f}$ ($n_{2\rm i}$ and $n_{2\rm f}$,
$n_{3\rm i}$ and $n_{3\rm f}$) are numbers of initial and final gluons
possessed by the first (second, third) triple-gluon vertex, respectively.}
\label{table2}
\begin{tabular*}{16cm}{@{\extracolsep{\fill}}c|c|c|c|c|c|c|c}

  \hline
  subclass & number of & $n_{1\rm i}$ & $n_{1\rm f}$ & $n_{2\rm i}$ & 
  $n_{2\rm f}$ & $n_{3\rm i}$ & $n_{3\rm f}$ \\
           & diagrams  &  &  &  &  &  &  \\
  \hline
  1 & 36 & 1 & 1 & 1 & 1 & 1 & 0 \\
  2 & 18 & 0 & 2 & 1 & 1 & 1 & 0 \\
  3 & 18 & 2 & 0 & 1 & 1 & 0 & 1 \\
  4 & 18 & 2 & 0 & 0 & 2 & 1 & 0 \\
  5 & 6  & 1 & 1 & 1 & 1 & 0 & 0 \\
  6 & 9  & 2 & 0 & 0 & 2 & 0 & 0 \\
  \hline
 \end{tabular*}
\end{table}

\begin{table}
\centering \caption{Number of diagrams in a subclass identified
by the four-gluon vertex and the selected triple-gluon vertex.
 $n_{{\rm i}4}$ and $n_{{\rm f}4}$ 
($n_{{\rm i}3}$ and $n_{{\rm f}3}$) are numbers of initial and final gluons
possessed by the 4-gluon vertex (the 3-gluon vertex), respectively.}
\label{table3}
\begin{tabular*}{16cm}{@{\extracolsep{\fill}}c|c|c|c|c|c}

  \hline
  subclass & number of & $n_{{\rm i}4}$ & $n_{{\rm f}4}$ & $n_{{\rm i}3}$ & 
  $n_{{\rm f}3}$ \\
           & diagrams  &  &  &  &  \\
  \hline
  1 & 18 & 1 & 2 & 1 & 1  \\
  2 & 18 & 2 & 1 & 1 & 1  \\
  3 & 9  & 1 & 2 & 2 & 0  \\
  4 & 9  & 2 & 1 & 0 & 2  \\
  5 & 3  & 0 & 3 & 2 & 0  \\
  6 & 3  & 3 & 0 & 0 & 2  \\
  7 & 18 & 1 & 1 & 1 & 1  \\
  8 & 9  & 0 & 2 & 2 & 0  \\
  9 & 9  & 2 & 0 & 0 & 2  \\
  \hline
 \end{tabular*}
\end{table}

\section{Transport equations and thermal states}
\medskip

We assume that quark-gluon matter consists of gluons, quarks and antiquarks
with up and down flavors and the two flavors (the quark and the antiquark as 
well) have the same amount. Let the distribution functions for the gluon, the
up-quark, the down-quark, the up-antiquark and the down-antiquark be
$f_{{\rm g}i}$, $f_{ui}$, $f_{di}$, $f_{\bar {u}i}$ and $f_{\bar {d}i}$, 
respectively, where $i$ labels the $i$th parton in scattering, and
$f_{ui}=f_{di}=f_{\bar {u}i}=f_{\bar {d}i}=f_{qi}$.
The transport equation for gluon matter is
\begin{eqnarray}
\frac {\partial f_{{\rm g}1}}{\partial t}
& + & \vec {\rm v}_1 \cdot \vec {\nabla}_{\vec {r}} f_{{\rm g}1} 
~~ = ~~ -\frac {1}{2E_1} \int \frac {d^3p_2}{(2\pi)^32E_2}
\frac {d^3p_3}{(2\pi)^32E_3} \frac {d^3p_4}{(2\pi)^32E_4}
(2\pi)^4 \delta^4(p_1+p_2-p_3-p_4)
         \nonumber    \\
& &
\times \left\{ \frac {{\rm g}_G}{2} 
\mid {\cal M}_{{\rm g}{\rm g} \to {\rm g}{\rm g}} \mid^2
[f_{{\rm g}1}f_{{\rm g}2}(1+f_{{\rm g}3})(1+f_{{\rm g}4})
-f_{{\rm g}3}f_{{\rm g}4}(1+f_{{\rm g}1})(1+f_{{\rm g}2})]  \right.
         \nonumber    \\
& &
+ {\rm g}_Q ( \mid {\cal M}_{{\rm g}u \to {\rm g}u} \mid^2
+ \mid {\cal M}_{{\rm g}d \to {\rm g}d} \mid^2
+ \mid {\cal M}_{{\rm g}\bar {u} \to {\rm g}\bar {u}} \mid^2
+ \mid {\cal M}_{{\rm g}\bar {d} \to {\rm g}\bar {d}} \mid^2 )
         \nonumber    \\
& &
\left.
\times [f_{{\rm g}1}f_{q2}(1+f_{{\rm g}3})(1-f_{q4})
-f_{{\rm g}3}f_{q4}(1+f_{{\rm g}1})(1-f_{q2})]
    \right\}
         \nonumber    \\
& &
-\frac {1}{2E_1} \int \frac {d^3p_2}{(2\pi)^32E_2}
\frac {d^3p_3}{(2\pi)^32E_3} \frac {d^3p_4}{(2\pi)^32E_4}
\frac {d^3p_5}{(2\pi)^32E_5} \frac {d^3p_6}{(2\pi)^32E_6}
         \nonumber    \\
& &
\times (2\pi)^4 \delta^4(p_1+p_2+p_3-p_4-p_5-p_6) 
\left\{ \frac {{\rm g}_G^2}{12} 
\mid {\cal M}_{{\rm g}{\rm g}{\rm g} \to {\rm  g}{\rm g}{\rm g}} \mid^2 \right.
         \nonumber    \\
& &
\times [f_{{\rm g}1}f_{{\rm g}2}f_{{\rm g}3}
(1+f_{{\rm g}4})(1+f_{{\rm g}5})(1+f_{{\rm g}6})-
f_{{\rm g}4}f_{{\rm g}5}f_{{\rm g}6}
(1+f_{{\rm g}1})(1+f_{{\rm g}2})(1+f_{{\rm g}3})]
         \nonumber    \\
& &
+ \frac {{\rm g}_G{\rm g}_Q}{2} 
( \mid {\cal M}_{{\rm g}{\rm g}u \to {\rm g}{\rm g}u} \mid^2
+\mid {\cal M}_{{\rm g}{\rm g}d \to {\rm g}{\rm g}d} \mid^2
+\mid {\cal M}_{{\rm g}{\rm g}\bar {u} \to {\rm g}{\rm g}\bar {u}} \mid^2 
+\mid {\cal M}_{{\rm g}{\rm g}\bar {d} \to {\rm g}{\rm g}\bar {d}} \mid^2 )
         \nonumber    \\
& &
\times [f_{{\rm g}1}f_{{\rm g}2}f_{q3}
(1+f_{{\rm g}4})(1+f_{{\rm g}5})(1-f_{q6})
-f_{{\rm g}4}f_{{\rm g}5}f_{q6}
(1+f_{{\rm g}1})(1+f_{{\rm g}2})(1-f_{q3})]
         \nonumber    \\
& &
+ {\rm g}_Q^2 [\frac {1}{4} \mid {\cal M}_{{\rm g}uu \to {\rm g}uu} \mid^2
+\frac {1}{2} ( \mid {\cal M}_{{\rm g}ud \to {\rm g}ud} \mid^2
              + \mid {\cal M}_{{\rm g}du \to {\rm g}du} \mid^2 )
+\frac {1}{4} \mid {\cal M}_{{\rm g}dd \to {\rm g}dd} \mid^2
         \nonumber    \\
& &
+ \mid {\cal M}_{{\rm g}u\bar {u} \to {\rm g}u\bar {u}} \mid^2
    + \mid {\cal M}_{{\rm g}u\bar {d} \to {\rm g}u\bar {d}} \mid^2
    + \mid {\cal M}_{{\rm g}d\bar {u} \to {\rm g}d\bar {u}} \mid^2
    + \mid {\cal M}_{{\rm g}d\bar {d} \to {\rm g}d\bar {d}} \mid^2
         \nonumber         \\
& &
+\frac {1}{4} \mid {\cal M}_{{\rm g}\bar {u}\bar {u}
                             \to {\rm g}\bar {u}\bar {u}} \mid^2
    +\frac {1}{2} ( \mid {\cal M}_{{\rm g}\bar {u}\bar {d}
                             \to {\rm g}\bar {u}\bar {d}} \mid^2 
    + \mid {\cal M}_{{\rm g}\bar {d}\bar {u} 
                             \to {\rm g}\bar {d}\bar {u}} \mid^2 )
+ \frac {1}{4} \mid {\cal M}_{{\rm g}\bar {d}\bar {d} 
                             \to {\rm g}\bar {d}\bar {d}} \mid^2 ]
         \nonumber    \\
& &
\left. \times [f_{{\rm g}1}f_{q2}f_{q3}(1+f_{{\rm g}4})(1-f_{q5})(1-f_{q6})
     -f_{{\rm g}4}f_{q5}f_{q6}(1+f_{{\rm g}1})(1-f_{q2})(1-f_{q3})] \right\} ,
         \nonumber    \\
\end{eqnarray}
and the transport equation for up-quark matter is
\begin{eqnarray}
\frac {\partial f_{q1}}{\partial t}
& + & \vec {\rm v}_1 \cdot \vec {\nabla}_{\vec {r}} f_{q1}
~~ = ~~ -\frac {1}{2E_1} \int \frac {d^3p_2}{(2\pi)^32E_2}
\frac {d^3p_3}{(2\pi)^32E_3} \frac {d^3p_4}{(2\pi)^32E_4}
(2\pi)^4 \delta^4(p_1+p_2-p_3-p_4)
         \nonumber    \\
& &
\times \left\{ {\rm g}_G \mid {\cal M}_{u{\rm g} \to u{\rm g}} \mid^2
[f_{q1}f_{{\rm g}2}(1-f_{q3})(1+f_{{\rm g}4})
-f_{q3}f_{{\rm g}4}(1-f_{q1})(1+f_{{\rm g}2})]  \right.
         \nonumber    \\
& &
+ {\rm g}_Q (\frac {1}{2} \mid {\cal M}_{uu \to uu} \mid^2
+ \mid {\cal M}_{ud \to ud} \mid^2 
+ \mid {\cal M}_{u\bar {u} \to u\bar {u}} \mid^2
+ \mid {\cal M}_{u\bar {d} \to u\bar {d}} \mid^2 )
         \nonumber    \\
& &
\times \left. [f_{q1}f_{q2}(1-f_{q3})(1-f_{q4})
-f_{q3}f_{q4}(1-f_{q1})(1-f_{q2})]   \right\}
         \nonumber    \\
& &
-\frac {1}{2E_1} \int \frac {d^3p_2}{(2\pi)^32E_2}
\frac {d^3p_3}{(2\pi)^32E_3} \frac {d^3p_4}{(2\pi)^32E_4}
\frac {d^3p_5}{(2\pi)^32E_5} \frac {d^3p_6}{(2\pi)^32E_6}
         \nonumber    \\
& &
\times (2\pi)^4 \delta^4(p_1+p_2+p_3-p_4-p_5-p_6)
\left\{ \frac {{\rm g}_G^2}{4} 
\mid {\cal M}_{u{\rm g}{\rm g} \to u{\rm g}{\rm g}} \mid^2  \right.
         \nonumber    \\
& &
\times [f_{q1}f_{{\rm g}2}f_{{\rm g}3}(1-f_{q4})
(1+f_{{\rm g}5})(1+f_{{\rm g}6})
-f_{q4}f_{{\rm g}5}f_{{\rm g}6}(1-f_{q1})(1+f_{{\rm g}2})(1+f_{{\rm g}3})]
          \nonumber     \\
& &
+{\rm g}_Q{\rm g}_G ( \frac {1}{2} 
\mid {\cal M}_{uu{\rm g} \to uu{\rm g}} \mid^2
+\mid {\cal M}_{ud{\rm g} \to ud{\rm g}} \mid^2
+\mid {\cal M}_{u\bar {u}{\rm g} \to u\bar {u}{\rm g}} \mid^2
+\mid {\cal M}_{u\bar {d}{\rm g} \to u\bar {d}{\rm g}} \mid^2 )
         \nonumber     \\
& &
\times [f_{q1}f_{q2}f_{{\rm g}3}(1-f_{q4})(1-f_{q5})(1+f_{{\rm g}6})
-f_{q4}f_{q5}f_{{\rm g}6}(1-f_{q1})(1-f_{q2})(1+f_{{\rm g}3})]
          \nonumber     \\
& &
+ {\rm g}_Q^2 [\frac {1}{12} \mid {\cal M}_{uuu \to uuu} \mid^2
+\frac {1}{4} ( \mid {\cal M}_{uud \to uud} \mid^2
              + \mid {\cal M}_{udu \to udu} \mid^2 )
+\frac {1}{4} \mid {\cal M}_{udd \to udd} \mid^2
         \nonumber    \\
& &
+\frac {1}{2} \mid {\cal M}_{uu\bar {u} \to uu\bar {u}} \mid^2
    +\frac {1}{2} \mid {\cal M}_{uu\bar {d} \to uu\bar {d}} \mid^2
            + \mid {\cal M}_{ud\bar {u} \to ud\bar {u}} \mid^2
            + \mid {\cal M}_{ud\bar {d} \to ud\bar {d}} \mid^2
         \nonumber         \\
& &
+\frac {1}{4} \mid {\cal M}_{u\bar {u}\bar {u}
                             \to u\bar {u}\bar {u}} \mid^2
    +\frac {1}{2} ( \mid {\cal M}_{u\bar {u}\bar {d}
                               \to u\bar {u}\bar {d}} \mid^2
+ \mid {\cal M}_{u\bar {d}\bar {u} \to u\bar {d}\bar {u}} \mid^2 )
+ \frac {1}{4}
  \mid {\cal M}_{u\bar {d}\bar {d} \to u\bar {d}\bar {d}} \mid^2 ]
         \nonumber    \\
& &
\times \left. [f_{q1}f_{q2}f_{q3}(1-f_{q4})(1-f_{q5})(1-f_{q6})
-f_{q4}f_{q5}f_{q6}(1-f_{q1})(1-f_{q2})(1-f_{q3})] \right\} ,
         \nonumber    \\
\end{eqnarray}
where $\rm \vec {v}_1$ is the parton velocity; ${\rm g}_G$ 
(${\rm g}_Q$) is the gluon (quark)
color-spin degeneracy factor; $p_i (i=1,\cdot \cdot \cdot,6)$ denote the
four-momenta of initial and final partons; $E_i$ is the energy component of
$p_i$. In equation (1) ${\cal M}_{gd\bar {d} \to gd\bar {d}}$ is the amplitude
for the elastic scattering of a gluon, a down-quark and a down-antiquark;
in equation
(2) ${\cal M}_{u\bar {d}\bar {u} \to u\bar {d}\bar {u}}$ is the amplitude for
the elastic scattering of an up-quark, a down-antiquark and an up-antiquark.
Other notations can be similarly understood. The transport
equations for down-quark matter, up-antiquark matter and down-antiquark matter
can be established just as the equation for up-quark matter.

We take the following replacement in the transport equations
\begin{displaymath}
f_{{\rm g}1}f_{{\rm g}2}(1+f_{{\rm g}3})(1+f_{{\rm g}4})
-f_{{\rm g}3}f_{{\rm g}4}(1+f_{{\rm g}1})(1+f_{{\rm g}2}) \to
f_{{\rm g}1}f_{{\rm g}2}-f_{{\rm g}3}f_{{\rm g}4},
\end{displaymath}
\begin{displaymath}
f_{{\rm g}1}f_{q2}(1+f_{{\rm g}3})(1-f_{q4})
-f_{{\rm g}3}f_{q4}(1+f_{{\rm g}1})(1-f_{q2}) \to
f_{{\rm g}1}f_{q2}-f_{{\rm g}3}f_{q4},
\end{displaymath}
\begin{displaymath}
f_{{\rm g}1}f_{{\rm g}2}f_{{\rm g}3}
(1+f_{{\rm g}4})(1+f_{{\rm g}5})(1+f_{{\rm g}6})
-f_{{\rm g}4}f_{{\rm g}5}f_{{\rm g}6}
(1+f_{{\rm g}1})(1+f_{{\rm g}2})(1+f_{{\rm g}3}) \to
f_{{\rm g}1}f_{{\rm g}2}f_{{\rm g}3}-f_{{\rm g}4}f_{{\rm g}5}f_{{\rm g}6},
\end{displaymath}
\begin{displaymath}
f_{{\rm g}1}f_{{\rm g}2}f_{q3}
(1+f_{{\rm g}4})(1+f_{{\rm g}5})(1-f_{q6})
-f_{{\rm g}4}f_{{\rm g}5}f_{q6}
(1+f_{{\rm g}1})(1+f_{{\rm g}2})(1-f_{q3}) \to
f_{{\rm g}1}f_{{\rm g}2}f_{q3}-f_{{\rm g}4}f_{{\rm g}5}f_{q6},
\end{displaymath}
\begin{displaymath}
f_{{\rm g}1}f_{q2}f_{q3}(1+f_{{\rm g}4})(1-f_{q5})(1-f_{q6})
-f_{{\rm g}4}f_{q5}f_{q6}(1+f_{{\rm g}1})(1-f_{q2})(1-f_{q3}) \to
f_{{\rm g}1}f_{q2}f_{q3}-f_{{\rm g}4}f_{q5}f_{q6},
\end{displaymath}
\begin{displaymath}
f_{q1}f_{{\rm g}2}(1-f_{q3})(1+f_{{\rm g}4})
-f_{q3}f_{{\rm g}4}(1-f_{q1})(1+f_{{\rm g}2}) \to
f_{q1}f_{{\rm g}2}-f_{q3}f_{{\rm g}4},
\end{displaymath}
\begin{displaymath}
f_{q1}f_{q2}(1-f_{q3})(1-f_{q4}) -f_{q3}f_{q4}(1-f_{q1})(1-f_{q2}) \to
f_{q1}f_{q2}-f_{q3}f_{q4},
\end{displaymath}
\begin{displaymath}
f_{q1}f_{{\rm g}2}f_{{\rm g}3}(1-f_{q4})(1+f_{{\rm g}5})(1+f_{{\rm g}6})
-f_{q4}f_{{\rm g}5}f_{{\rm g}6}(1-f_{q1})(1+f_{{\rm g}2})(1+f_{{\rm g}3}) \to
f_{q1}f_{{\rm g}2}f_{{\rm g}3}-f_{q4}f_{{\rm g}5}f_{{\rm g}6},
\end{displaymath}
\begin{displaymath}
f_{q1}f_{q2}f_{{\rm g}3}(1-f_{q4})(1-f_{q5})(1+f_{{\rm g}6})
-f_{q4}f_{q5}f_{{\rm g}6}(1-f_{q1})(1-f_{q2})(1+f_{{\rm g}3}) \to
f_{q1}f_{q2}f_{{\rm g}3}-f_{q4}f_{q5}f_{{\rm g}6},
\end{displaymath}
\begin{displaymath}
f_{q1}f_{q2}f_{q3}(1-f_{q4})(1-f_{q5})(1-f_{q6})
-f_{q4}f_{q5}f_{q6}(1-f_{q1})(1-f_{q2})(1-f_{q3}) \to
f_{q1}f_{q2}f_{q3}-f_{q4}f_{q5}f_{q6}.
\end{displaymath}
Then, any Monte Carlo method can be applied to solve the equations. We generate
1500 gluons, 250 up-quarks, 250 down-quarks, 250 up-antiquarks and 250 
down-antiquarks within $-0.3<z<0.3$ fm in the longitudinal direction and 
$r<6.4$ fm in the transverse direction from HIJING for
central Au-Au collisions at $\sqrt {s_{NN}}=200$ GeV. The number of
partons moving in the longitudinal direction is much larger than the number of
partons moving in the transverse direction. The anisotropically distributed
partons undergo elastic 2-to-2 scattering \cite{cutler,combridge} and
elastic 3-to-3 scattering to eventually become isotropic in momentum space
(i.e. become a thermal state) at a time. But from gluon matter to quark matter
the time is different, i.e. the thermalization time of gluon matter differs
from the thermalization time of quark matter. At $t=0.52$ fm/$c$ the gluon
momentum distribution meets the J\"uttner distribution
\begin{equation}
f_g(\vec {p})=\frac {\lambda_g}{{\rm e}^{\mid \vec {p}\mid/T}-\lambda_g},
\end{equation}
with the temperature $T=0.52$ GeV and the gluon fugacity $\lambda_g =0.328$
that label the thermal state of gluon matter. The corresponding thermalization
time is 0.32 fm/$c$. At $t=0.86$ fm/$c$ the quark momentum distribution meets
the J\"uttner distribution
\begin{equation}
f_q(\vec {p})=\frac {\lambda_q}{{\rm e}^{\mid \vec {p}\mid/T}+\lambda_q},
\end{equation}
with $T=0.46$ GeV and the quark fugacity $\lambda_q=0.143$ that label the
thermal state of quark matter. The corresponding thermalization time is 0.66
fm/$c$. The thermalization time of quark matter is larger than the one of gluon
matter.

\section{Conclusions}
\medskip
The elastic 2-to-2 scattering and the elastic 3-to-3 scattering lead to the
rapid creation of temperature or the early thermalization of quark-gluon 
matter. Solutions of the transport equations offer the thermalization times and
the thermal states of gluon matter and quark matter. Gluon matter thermalizes
faster than quark matter.

\ack
This work was supported by the National Natural Science Foundation of China
under grant no 11175111.

\end{document}